\documentclass[envcountsame]{llncs}
\usepackage{amsmath}
\usepackage{amssymb}
\usepackage{gastex}
\usepackage{graphicx}

\newenvironment{Theorem}{\begin{theorem}}{\end{theorem}}
\newenvironment{Proposition}{\begin{proposition}}{\end{proposition}}
\newenvironment{Lemma}{\begin{lemma}}{\end{lemma}}
\newenvironment{Corollary}{\begin{corollary}}{\end{corollary}}

\newenvironment{Proof}{\begin{proof}}{\qed\end{proof}}
\newenvironment{Example}{\begin{example}\rm}{\qed\end{example}}

\newcommand{\two}{\mathbb{F}_2}
\newcommand{\sub}[1]{[ #1 ]}
\newcommand{\xor}{\oplus}
\newcommand{\pair}[2]{\{{#1},{#2}\}}
\newcommand{\vertexrem}{\setminus}

\renewcommand{\emptyset}{\varnothing}

\begin{document}

\title{Nullity Invariance for Pivot\\ and the Interlace Polynomial}
\author{Robert Brijder\thanks{corresponding author: \email{rbrijder@liacs.nl}} \and Hendrik Jan Hoogeboom}

\institute{Leiden Institute of Advanced Computer Science,\\
Leiden University, The Netherlands}

\maketitle
\begin{abstract}
We show that the effect of principal pivot transform on the nullity
values of the principal submatrices of a given (square) matrix is
described by the symmetric difference operator (for sets). We
consider its consequences for graphs, and in particular generalize
the recursive relation of the interlace polynomial and simplify its
proof.
\end{abstract}

\section{Introduction}
Principal pivot transform (PPT, or simply pivot) is a matrix
transformation operation capable of partially (component-wise)
inverting a given matrix. PPT is originally motivated by the
well-known linear complementarity problem \cite{tucker1960}, and is
applied in many other settings such as mathematical programming and
numerical analysis, see \cite{Tsatsomeros2000151} for an overview.

A natural restriction of pivot is to graphs (with possibly loops),
i.e., symmetric matrices over $\two$. For graphs, each pivot
operation can be decomposed into a sequence of elementary pivots.
There are two types of elementary pivot operations, (frequently)
called local complementation and edge complementation. These two
graph operations are also (in fact, originally) defined for simple
graphs. The operations are similar for graphs and simple graphs,
however, for simple graphs, applicability is less restrictive. Local
and edge complementation for simple graphs, introduced in
\cite{kotzig1968} and \cite{bouchet1988} respectively, were
originally motivated by the study of Euler circuits in 4-regular
graphs and by the study of circle graphs (also called overlap
graphs) as they model natural transformations of the underlying
circle segments. Many other applications domains for these
operations have since appeared, e.g., quantum computing
\cite{PhysRevA.69.022316}, the formal theory of gene assembly in
ciliates \cite{GeneAssemblyBook} (a research area within
computational biology), and the study of interlace polynomials,
initiated in \cite{Arratia_InterlaceP_SODA}. In many contexts where
local and edge complementation have been used, PPT has originally
appeared in disguise (we briefly discuss some examples in the
paper).

In this paper we show that the pivot operator on matrices $A$ (over
some field) and the symmetric difference operator on sets $Y$ have
an equivalent effect w.r.t. the nullity value of the principal
submatrices $A\sub{Y}$ of $A$. We subsequently show that this
nullity invariant can be formulated in terms of (a sequence of) set
systems. Furthermore we discuss its consequences for pivot on graphs
and in particular apply it to the interlace polynomial. It was shown
in \cite{ArratiaBS04} that the interlace polynomial, which is
defined for graphs, fulfills a characteristic recursive relation. We
generalize the notion of interlace polynomial and its recursive
relation to square matrices in general. In this way, we simplify the
proof of the (original) recursive relation for interlace polynomials
of graphs. Also, in Section~\ref{sec:background}, we recall a
motivation of pivot applied to overlap graphs, and relate it to the
nullity invariant.

\section{Notation and Terminology}
A \emph{set system} (over $V$) is a tuple $M = (V,D)$ with $V$ a
finite set, called the \emph{domain} of $M$, and $D \subseteq 2^V$ a
family of subsets of $V$. To simplify notation we often write $X \in
M$ to denote $X \in D$. Moreover, we often simply write $V$ to
denote the domain of the set system under consideration. We denote
by $\xor$ the logical exclusive-or (i.e., addition in $\two$), and
we carry this operator over to sets: for sets $A, B \subseteq V$, $A
\xor B$ is the set defined by $x \in A \xor B$ iff $(x \in A) \xor
(x \in B)$ for $x \in V$. For sets, the $\xor$ operator is called
symmetric difference.

We consider matrices and vectors indexed by a finite set $V$. For a
vector $v$ indexed by $V$, we denote the element of $v$
corresponding to $i \in V$ by $v[i]$. Also, we denote the nullity
(dimension of the null space) and the determinant of a matrix $A$ by
$n(A)$ and $\det(A)$ respectively. For $X \subseteq V$, the
principal submatrix of $A$ w.r.t. $X$ is denoted by $A\sub{X}$.

We consider undirected graphs without parallel edges, however we do
allow loops. Hence a graph $G=(V,E)$ can be considered a symmetric
$V \times V$-matrix $A = \left(a_{u,v}\right)$ over $\two$ (the
field having two elements): for $u \in V$, $\{u\} \in E$ (i.e., $u$
has a loop in $G$) iff $a_{u,u} = 1$, and for $u,v \in V$ with $u
\not=v$, $\{u,v\} \in E$ iff $a_{u,v} = 1$. We denote the set of
edges of $G$ by $E(G)$. We often make no distinction between $G$ and
its matrix representation $A$. Thus, e.g., we write $n(G) = n(A)$,
and, for $X \subseteq V$, $G\sub{X} = A\sub{X}$, which consequently
is the subgraph of $G$ induced by $X$. Note that as $G$ is
represented by a matrix $A$ over $\two$, $n(G)$ is computed over
$\two$. Also, for $Y \subseteq V$, we define $G \vertexrem Y = G
\sub{V\setminus Y}$. In case $Y = \{v\}$ is a singleton, to simplify
notation, we also write $G \vertexrem Y = G \vertexrem v$. Similar
as for set systems, we often write $V$ to denote the vertex set of
the graph under consideration.

\section{Background: Nullity and Counting Closed Walks}
\label{sec:background} In this section we briefly and informally
discuss an application of principal pivot transform where nullity
plays an important role. In \cite{DBLP:journals/jct/CohnL72} a first
connection between counting cycles and the nullity of a suitable
matrix was established. It is shown in that paper that the number of
cycles obtained as the result of applying disjoint transpositions to
a cyclic permutation is described by the nullity of a corresponding
``interlace matrix''.

It has been recognized in \cite{LT/BinNullity/09} that the result of
\cite{DBLP:journals/jct/CohnL72} has an interpretation in terms of
2-in, 2-out digraphs (i.e., directed graphs with 2 incoming and 2
outgoing edges for each vertex), linking it to the interlace
polynomial \cite{Arratia2004199}.
%
%
We discuss now this interpretation in terms of 2-in, 2-out digraphs
and subsequently show the connection to the pivot operation.

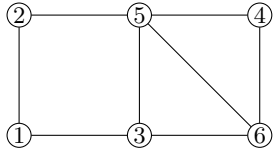
\begin{figure}[t]
\begin{center}
\unitlength 0.4mm%
\begin{picture}(80,40)
  \gasset{AHnb=0,Nw=1.5,Nh=1.5,Nframe=n,Nfill=y}
  \gasset{AHnb=0,Nw=8,Nh=8,Nframe=y,Nfill=n}
  \node(1)(0,0){$1$}
  \node(2)(0,40){$2$}
  \node(3)(40,0){$3$}
  \node(5)(40,40){$5$}
  \node(6)(80,0){$6$}
  \node(4)(80,40){$4$}
  \drawedge(1,2){}
  \drawedge(1,3){}
  \drawedge(2,5){}
  \drawedge(3,5){}
  \drawedge(5,4){}
  \drawedge(5,6){}
  \drawedge(3,6){}
  \drawedge(4,6){}
\end{picture}
\end{center} \caption{The overlap graph of $s = 146543625123$.}
\label{fig:overlap_graph}
\end{figure}

Let $V = \{1,2,3,4,5,6\}$ be an alphabet and let $s = 146543625123$
be a double occurrence string (i.e., each letter of the string
occurs precisely twice) over $V$. The \emph{overlap graph} $O_s$
corresponding to $s$ has $V$ as the set of vertices and an edge
$\{u,v\}$ precisely when $u$ and $v$ overlap: the vertices $u$ and
$v$ appear either in order $u,v,u,v$ or in order $v,u,v,u$ in $s$.
The overlap graph $O_s$ is given in Figure~\ref{fig:overlap_graph}.
One may verify that the nullity of $O_s$ is $n(O_s) = 0$. Consider
now the subgraph $O_s[X]$ of $O_s$ induced by $X = \{3,4,5,6\}$.
Then it can be verified that $n(O_s[X]) = 2$.

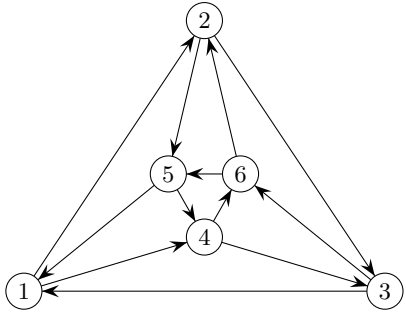
\begin{figure}[t]
\begin{center}
\unitlength 0.6mm%
\begin{picture}(80,60)
  \gasset{AHLength=4,Nw=8,Nh=8,Nframe=y,Nfill=n}
  \node(1)(0,0){$1$}
  \node(2)(40,60){$2$}
  \node(3)(80,0){$3$}
  \node(4)(40,12){$4$}
  \node(5)(32,26){$5$}
  \node(6)(48,26){$6$}
  \drawedge(1,2){}
  \drawedge(3,1){}
  \drawedge(2,3){}
  \drawedge(5,4){}
  \drawedge(4,6){}
  \drawedge(6,5){}
  \drawedge(1,4){}
  \drawedge(5,1){}
  \drawedge(2,5){}
  \drawedge(6,2){}
  \drawedge(3,6){}
  \drawedge(4,3){}
\end{picture}
\end{center} \caption{A 2-in, 2-out digraph.}
\label{fig:4_reg_graph}
\end{figure}

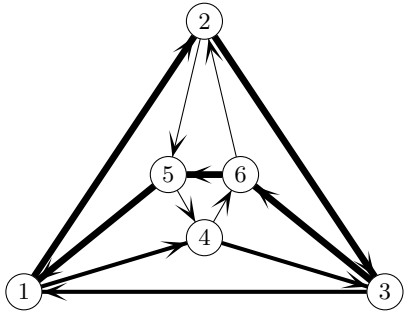
\begin{figure}[t]
\begin{center}
\unitlength 0.6mm%
\begin{picture}(80,60)
  \gasset{AHLength=6,Nw=8,Nh=8,Nframe=y,Nfill=n}
  \node(1)(0,0){$1$}
  \node(2)(40,60){$2$}
  \node(3)(80,0){$3$}
  \node(4)(40,12){$4$}
  \node(5)(32,26){$5$}
  \node(6)(48,26){$6$}
  \drawedge[linewidth=1.5](1,2){}
  \drawedge[linewidth=0.9](3,1){}
  \drawedge[linewidth=1.5](2,3){}
  \drawedge(5,4){}
  \drawedge(4,6){}
  \drawedge[linewidth=1.5](6,5){}
  \drawedge[linewidth=0.9](1,4){}
  \drawedge[linewidth=1.5](5,1){}
  \drawedge(2,5){}
  \drawedge(6,2){}
  \drawedge[linewidth=1.5](3,6){}
  \drawedge[linewidth=0.9](4,3){}
\end{picture}
\end{center} \caption{Partition of the edges of a 2-in, 2-out digraph into three closed walks.}
\label{fig:4_reg_graph_closed_walks}
\end{figure}

We discuss now the link with 2-in, 2-out digraphs (only in this
section we consider digraphs). Let $G$ be the 2-in, 2-out digraph of
Figure~\ref{fig:4_reg_graph} with $V = \{1,2,3,4,5,6\}$ as the set
of vertices. Although our example graph does not have parallel
edges, there is no objection to consider such ``2-in, 2-out
multidigraphs''. Notice that the double occurrence string $s =
146543625123$ considered earlier corresponds to an Euler circuit $C$
of $G$. We now consider partitions $P$ of the edges of $G$ into
closed walks (i.e., cycles where repeated vertices are allowed).
Note that there are $2^{|V|}$ such partitions: if in a walk passing
through vertex $v$ we go \emph{from} incoming edge $e$ of $v$
\emph{to} outgoing edge $e'$ of $v$, then necessarily we also walk
in $P$ from the other incoming edge of $v$ to the other outgoing
edge of $v$. Hence for each vertex there are two ``routes''. Let $P$
now be the the partition of the edges of $G$ into $3$ closed walks
as indicated by Figure~\ref{fig:4_reg_graph_closed_walks} using
three types of line thicknesses. Then $P$ follows the same route as
the Euler circuit (corresponding to) $s$ in vertices $\{1,2\}$,
while in the other vertices $X = \{3,4,5,6\}$ it follows the other
route. We say that $P$ is \emph{induced} by $X$ in $s$.

Theorem~1 in \cite{DBLP:journals/jct/CohnL72} now states (applying
it to the context of 2-in, 2-out digraphs) that the number of closed
walks of a partition $P$ of edges induced by $X$ in $s$ is
$n(O_s[X])+1$. In our case we have indeed $|P| = 3$ and
$n(O_s[X])=2$.

The pivot operation, which is recalled in the next section, has the
property that it can map $O_{s_1}$ into $O_{s_2}$ for any two double
occurrence strings $s_1$ and $s_2$ that correspond to Euler circuits
of a 2-in, 2-out digraph $G$, see, e.g., the survey section of
\cite{DBLP:journals/ejc/Bouchet01}. For example, the partition of
edges induced by $\{1,3\}$ in $s$ corresponds to a single closed
walk which may be described by the double occurrence string $s' =
123625146543$. It then holds that $O_{s'}$ is obtained from $O_{s}$
by pivot on $\{1,3\}$, denoted by $O_{s'} = O_{s}*\{1,3\}$. We
notice that the partition induced by $\{1,3\} \xor \{3,4,5,6\} =
\{1,4,5,6\}$ in $s'$ is equal to the partition $P$ induced by
$\{3,4,5,6\}$ in $s$ depicted in
Figure~\ref{fig:4_reg_graph_closed_walks}. Hence $n(O_{s}*Y[Y \xor
X]) = n(O_s[X])$ for $X = \{3,4,5,6\}$ and $Y = \{1,3\}$. In
Theorem~\ref{thm:nullity_invariant} below we prove this property for
arbitrary $X$ and $Y$ and for arbitrary square matrices (over some
field) instead of restricting to overlap graphs $O_{s}$.

\section{Pivot} \label{sec:def_pivots}
In this section we recall principal pivot transform (pivot for
short) for square matrices over an arbitrary field in general, see
also \cite{Tsatsomeros2000151}.

Let $A$ be a $V \times V$-matrix (over an arbitrary field), and let
$X \subseteq V$ be such that the corresponding principal submatrix
$A\sub{X}$ is nonsingular, i.e., $\det A\sub{X} \neq 0$. The
\emph{pivot} of $A$ on $X$, denoted by $A*X$, is defined as follows.
If $P = A\sub{X}$ and $A = \left(
\begin{array}{c|c}
P & Q \\
\hline R & S
\end{array}
\right)$, then
\begin{eqnarray}\label{pivot_def}
A*X = \left(
\begin{array}{c|c}
P^{-1} & -P^{-1} Q \\
\hline R P^{-1} & S - R P^{-1} Q
\end{array}
\right).
\end{eqnarray}
Matrix $S - R P^{-1} Q$ is called the \emph{Schur complement} of $P$
in $A$.

The pivot can be considered a partial inverse, as $A$ and $A*X$ are
related by the following equality, where the vectors $x_1$ and $y_1$
correspond to the elements of $X$. This equality is characteristic
as it is sufficient to define the pivot operation, see
\cite[Theorem~3.1]{Tsatsomeros2000151}.
\begin{eqnarray} \label{pivot_def_reverse} A \left(
\begin{array}{c} x_1 \\ x_2 \end{array} \right) = \left(\begin{array}{c} y_1 \\ y_2 \end{array} \right) \mbox{ iff } A*X \left(
\begin{array}{c} y_1 \\ x_2 \end{array} \right) = \left(\begin{array}{c} x_1 \\ y_2 \end{array}
\right) \end{eqnarray} Note that if $\det A \not= 0$, then $A * V =
A^{-1}$. Also note by Equation~(\ref{pivot_def_reverse}) that the
pivot operation is an involution (operation of order $2$), and more
generally, if $(A*X)*Y$ is defined, then it is equal to $A*(X \xor
Y)$.

\section{Nullity Invariant} \label{sec:nullity_invar}
It is well known that any Schur complement in a matrix $A$ has the
same nullity as $A$ itself, see, e.g.,
\cite[Section~6.0.1]{SchurBook2005}. See moreover
\cite[Chapter~0]{SchurBook2005} for a detailed historical account of
the Schur complement. We can rephrase the nullity property of the
Schur complement in terms of pivot as follows.
\begin{Proposition}[Nullity of Schur complement] \label{prop:Schur}
Let $A$ be a $V \times V$-matrix, and let $X \subseteq V$ such that
$A\sub{X}$ is nonsingular. Then $n(A*X[V \backslash X]) = n(A[V])$.
\end{Proposition}

The following result is known from \cite{tucker1960} (see also
\cite[Theorem~4.1.2]{cottle1992}).

\begin{Proposition}\label{prop:tucker}
Let $A$ be a $V\times V$-matrix, and let $X\subseteq V$ be such that
$A\sub{X}$ is nonsingular. Then, for $Y \subseteq V$,
$$
\det (A*X)\sub{Y} = \det A\sub{X \xor Y} / \det A\sub{X}.
$$
\end{Proposition}

As a consequence of Proposition~\ref{prop:tucker} we have the
following result.

\begin{Corollary}\label{cor:tucker}
Let $A$ be a $V\times V$-matrix, and let $X\subseteq V$ be such that
$A\sub{X}$ is nonsingular. Then, for $Y \subseteq V$, $(A*X)\sub{Y}$
is nonsingular iff $A\sub{X \xor Y}$ is nonsingular.
\end{Corollary}

We will now combine and generalize Proposition~\ref{prop:Schur} and
Corollary~\ref{cor:tucker} to obtain
Theorem~\ref{thm:nullity_invariant} below.

We denote by $A \sharp X$ the matrix obtained from $A$ by replacing
every row $v_x^T$ of $A$ belonging to $x \in V \setminus X$ by
$i_x^T$ where $i_x$ is the vector having value $1$ at element $x$
and $0$ elsewhere.

\begin{Lemma} \label{lem:nullity_principal_submatrix}
Let $A$ be a $V \times V$-matrix and $X \subseteq V$. Then $n(A
\sharp X) = n(A\sub{X})$.
\end{Lemma}
\begin{Proof}
By rearranging the elements of $V$, $A$ is of the following form
$\left(\begin{array}{c|c}
P & Q \\
\hline R & S
\end{array}\right)$
where $A\sub{X} = P$. Now $A \sharp X$ is $\left(\begin{array}{c|c}
P & Q \\
\hline 0 & I
\end{array}\right)$
where $I$ is the identity matrix of suitable size. We have $n(P) =
n(A \sharp X)$.
\end{Proof}

We are now ready to prove the following result, which we refer to as
the \emph{nullity invariant}.
\begin{Theorem}\label{thm:nullity_invariant}
Let $A$ be a $V\times V$-matrix, and let $X\subseteq V$ be such that
$A\sub{X}$ is nonsingular. Then, for $Y \subseteq V$,
$n((A*X)\sub{Y}) = n(A\sub{X \xor Y})$.
\end{Theorem}
\begin{Proof}
We follow the same line of reasoning as the proof of
Parsons\cite{ParsonsTDP70} of Proposition~\ref{prop:tucker} (see
also \cite[Theorem 4.1.1]{cottle1992}). Let $Ax = y$. Then
$$
((A \sharp X)x)[i] = \begin{cases} y[i] & \mbox{if } i \in X,\\
x[i] & \mbox{otherwise}.
\end{cases}
$$
As, by Equation~(\ref{pivot_def_reverse}),
$$
((A*X)(A \sharp X)x)[i] = \begin{cases} x[i] & \mbox{if } i \in X,\\
y[i] & \mbox{otherwise},
\end{cases}
$$
we have, by considering each of the four cases depending on whether
or not $i$ in $X$ and $i$ in $Y$ separately,
$$
(((A*X) \sharp Y)(A \sharp X)x)[i] = \begin{cases} y[i] & \mbox{if } i \in X \xor Y,\\
x[i] & \mbox{otherwise}.
\end{cases}
$$
Thus we have $((A*X) \sharp Y) (A \sharp X) = A \sharp (X \xor Y)$.
By Lemma~\ref{lem:nullity_principal_submatrix}, $n(A \sharp X) =
n(A\sub{X}) = 0$, and therefore $A \sharp X$ is invertible.
Therefore $n((A*X) \sharp Y) = n(A \sharp (X \xor Y))$, and the
result follows by Lemma~\ref{lem:nullity_principal_submatrix}.
\end{Proof}

By Theorem~\ref{thm:nullity_invariant}, we see that the pivot
operator $*X$ on matrices and the symmetric difference operator
$\xor X$ on sets have an equivalent effect on the nullity values of
principal submatrices.

Note that Theorem~\ref{thm:nullity_invariant} generalizes
Corollary~\ref{cor:tucker} as a matrix is nonsingular iff the
nullity of that matrix is $0$ (the empty matrix is nonsingular by
convention). One can immediately see that
Theorem~\ref{thm:nullity_invariant} generalizes
Proposition~\ref{prop:Schur}.

Also note that by replacing $Y$ by $V \setminus Y$ in
Theorem~\ref{thm:nullity_invariant}, we also have, equivalently,
$n((A*X)\sub{X \xor Y}) = n(A\sub{Y})$.

The ``Nullity Theorem'' \cite[Theorem~2]{Fiedler1986}, restricted to
\emph{square} principal submatrices, states that if $A$ is an
invertible $V\times V$-matrix, then, for $Y \subseteq V$,
$n(A^{-1}\sub{Y}) = n(A\sub{V \setminus Y})$. Note that this is
implied by Theorem~\ref{thm:nullity_invariant} as $A * V = A^{-1}$.

\begin{Example} \label{ex:matrix_nullity_invar}
Let $V = \{a,b,c\}$ and let $A$ be the $V \times V$-matrix
$\left(\begin{array}{ccc}
1 & 2 & 5\\
1 & 4 & 2\\
3 & 2 & 1
\end{array}
\right)$ over $\mathbb{Q}$ where the columns and rows are indexed by
$a,b,c$ respectively. We see that $A\sub{\{b,c\}} =
\left(\begin{array}{cc} 4 & 2\\ 2 & 1
\end{array}
\right)$ and therefore $n(A\sub{\{b,c\}})=1$. Moreover, for $X =
\{a,b\}$, the columns of $A\sub{X}$ are independent and thus $\det
A\sub{X} \not= 0$. We have therefore that $A*X$ is defined, and it
is given below.
$$
A*X = \left(\begin{array}{ccc}
2 & -1 & -8\\
-\frac{1}{2} & \frac{1}{2} & \frac{3}{2}\\
5 & -2 & -20
\end{array}
\right)
$$
By Theorem~\ref{thm:nullity_invariant}, we have $n(A\sub{\{b,c\}}) =
n(A*X[X \xor \{b,c\}]) = n(A*X[\{a,c\}])$. Therefore
$n(A*X[\{a,c\}]) = 1$. This can easily be verified given
$A*X[\{a,c\}]= \left(\begin{array}{cc}
2 & -8\\
5 & -20
\end{array}
\right)
$
\end{Example}

It is easy to verify from the definition of pivot that $A*X$ is
skew-symmetric whenever $A$ is. In particular, if $G$ is a graph
(i.e., a symmetric matrix over $\two$), then $G*X$ is also a graph.
For graphs, all matrix computations, including the determinant, will
be over $\two$.

\newcommand{\pivotorbitmacro}{
  \gasset{AHLength=0,Nw=8,Nh=8,Nframe=y,Nfill=n}
  \node(1)(0,30){$1$}
  \node(2)(30,30){$2$}
  \node(3)(0,0){$3$}
  \node(4)(30,0){$4$}
}

\begin{figure}[t]
\begin{center}
\begin{picture}(90,30)(0,0)
\unitlength 1mm%
\node[Nw=25,Nh=25,Nframe=n](I)(15,15){%
\unitlength 0.4mm%
\begin{picture}(30,30)
\pivotorbitmacro
  \drawedge(2,3){}
  \drawedge(2,4){}
  \drawedge(1,3){}
  \drawedge(3,4){}
  \drawloop[loopangle=90,loopdiam=7](2){}
  \drawloop[loopangle=-90,loopdiam=7](3){}
\end{picture}
}
\node[Nw=25,Nh=25,Nframe=n](II)(75,15){%
\unitlength 0.4mm%
\begin{picture}(30,30)
\pivotorbitmacro
  \drawedge(2,1){}
  \drawedge(2,4){}
  \drawedge(1,3){}
  \drawloop[loopangle=90,loopdiam=7](2){}
  \drawloop[loopangle=-90,loopdiam=7](4){}
\end{picture}
} \drawedge[AHnb=1,ATnb=1](I,II){$*\{1,2,3\}$}
\end{picture}
\end{center} \caption{Graphs $G$ and $G*X$ of Example~\ref{ex:graph_nullity_invar}.}
\label{fig:ex_graph_nullity_invar}
\end{figure}
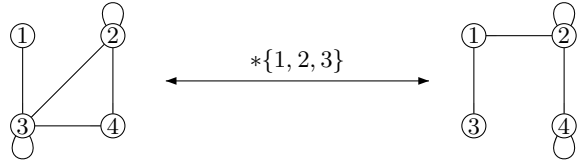

\begin{Example} \label{ex:graph_nullity_invar}
Let $G$ be the graph given on the left-hand side of
Figure~\ref{fig:ex_graph_nullity_invar}. Let $X = \{1,2,3\}$. Then
the $X \times X$-matrix belonging to $G\sub{X}$ is
$\left(\begin{array}{ccccc}
0 & 0 & 1\\
0 & 1 & 1\\
1 & 1 & 1
\end{array}
\right)$ where the columns and rows represent vertices $1,2,3$,
respectively. We see that the columns of $G\sub{X}$ are independent
(over $\two$) and therefore $\det G\sub{X} = 1$. Consequently $G*X$
is defined and the graph is given on the right-hand side of
Figure~\ref{fig:ex_graph_nullity_invar}. Let now $Y = \{1,4\}$. We
see that $G\sub{Y}$ is a discrete graph (i.e., the graph has no
edges). Therefore $n(G\sub{Y}) = 2$. Now by
Theorem~\ref{thm:nullity_invariant}, we have $n(G\sub{Y}) = n(G*X[X
\xor Y]) = n(G*X[\{2,3,4\}])$. One may verify that removing vertex
$1$ from $G*X$ indeed obtains a graph of nullity $2$.
\end{Example}

\section{Set Systems} \label{sec_seq_set_systems}
Let $A$ be a $V\times V$-matrix. Let $\mathcal{M}_A = (V,D)$ be the
set system with $X \in D$ iff $A\sub{X}$ is nonsingular. Set system
$\mathcal{M}_A$ turns out to fulfill a specific exchange axiom if
$A$ is (skew-)symmetric, making it in this case a delta-matroid
\cite{bouchet1987} (we will not recall its definition here as we do
not use this notion explicitly).

Let $M = (V,D)$ be a set system. We define for $X \subseteq V$, the
\emph{pivot} (often called \emph{twist}) of $M$ on $X$, denoted
$M*X$, by $(V,D*X)$ where $D*X = \{Y \xor X \mid Y \in M\}$. By
Corollary~\ref{cor:tucker} it is easy to verify, see
\cite{Geelen97}, that the operations of pivot on set systems and
matrices match, i.e., $\mathcal{M}_{A}*X = \mathcal{M}_{A*X}$ if the
right-hand side is defined (i.e., if $X \in \mathcal{M}_{A}$).
%

Theorem~\ref{thm:nullity_invariant} allows now for a generalization
of this result from the set system $\mathcal{M}_{A}$ of nullity $0$
to a ``sequence of set systems'' $\mathcal{P}_A$ for each possible
nullity $i$. We formalize this as follows.

For a finite set $V$, we call a sequence $P = (P_0,P_1,\ldots,P_n)$
with $n = |V|$ and $P_i \subseteq V$ for all $i \in \{0,\ldots,n\}$
a \emph{partition sequence} (over $V$) if the nonempty $P_i$'s form
a partition of $2^V$. Regarding $P$ as a vector indexed by
$\{0,\ldots,n\}$, we denote $P_i$ by $P[i]$. Moreover, we define for
partition sequence $P$ and $X \subseteq V$, the \emph{pivot} of $P$
on $X$, denoted by $P*X$, to be the partition sequence
$(P_0*X,P_1*X,\ldots,P_n*X)$. Also, we call the vector
$(|P_0|,|P_1|,\ldots,|P_n|)$ of dimension $n+1$, denoted by $\|P\|$,
the \emph{norm} of $P$. Clearly, $\|P\| = \|P*X\|$, i.e., the norm
of $P$ is invariant under pivot.

For a $V \times V$-matrix $A$ we denote by $\mathcal{P}_A$ the
partition sequence over $V$ where $X \in \mathcal{P}_A[i]$ iff
$n(A\sub{X}) = i$. As nullity $0$ corresponds to a non-zero
determinant (this holds also for $\emptyset$ as $\det
A\sub{\emptyset} = 1$ by convention), we have $\mathcal{M}_A =
(V,\mathcal{P}_A[0])$.

We now have the following consequence of
Theorem~\ref{thm:nullity_invariant}. Note that $X \in
\mathcal{P}_A[0]$ iff $A*X$ is defined.
\begin{Theorem} \label{thm:twist_gen_matroid}
Let $A$ be a $V\times V$-matrix, and $X \in \mathcal{P}_A[0]$. Then
$\mathcal{P}_{A*X} = \mathcal{P}_A*X$.
\end{Theorem}
\begin{Proof}
By Theorem~\ref{thm:nullity_invariant} we have for all
$i\in\{0,\ldots,n\}$, $Y \in \mathcal{P}_{A*X}[i]$ iff $n((A*X)[Y])
= i$ iff $n(A[X \xor Y]) = i$ iff $X \xor Y \in \mathcal{P}_{A}[i]$
iff $Y \in \mathcal{P}_{A}[i]*X$.
\end{Proof}
Since the norm of a partition sequence is invariant under pivot, we
have by Theorem~\ref{thm:twist_gen_matroid}, $\|\mathcal{P}_A\| =
\|\mathcal{P}_{A*X}\|$. Therefore, for each $i \in \{0,\ldots,n\}$,
the number of principal submatrices of $A$ of nullity $i$ is equal
to the number of principal submatrices of $A*X$ of nullity $i$.

For $X \subseteq V$, it it is easy to see that $\mathcal{P}_{A[X]}$
is obtained from $\mathcal{P}_A$ by removing all $Y \in
\mathcal{P}_A[i]$ containing at least one element outside $X$:
$\mathcal{P}_{A[X]}[i] = \{Z \subseteq X \mid Z \in
\mathcal{P}_A[i]\}$ for all $i \in \{0,\ldots,|X|\}$.

\begin{Example}
For matrix $A$ from Example~\ref{ex:matrix_nullity_invar}, we have
$\mathcal{P}_A = (P_0,P_1,P_2,P_3)$ with $P_0 = 2^V\setminus
\{\{b,c\}\}$, $P_1 = \{\{b,c\}\}$, and $P_2 = P_3 = \emptyset$.
\end{Example}

\begin{Example}
For graph $G$ from Example~\ref{ex:graph_nullity_invar}, depicted on
the left-hand side of Figure~\ref{fig:ex_graph_nullity_invar}, we
have $\mathcal{P}_G = (P_0,P_1,P_2,P_3,P_4)$ with
\begin{eqnarray*}
P_0 &=& \{\emptyset, \{2\}, \{3\}, \{1,3\}, \{2,4\}, \{3,4\},
\{1,2,3\}, \{1,2,3,4\}\}, \\
P_1 &=& \{\{1\}, \{4\}, \{1,2\}, \{2,3\}, \{1,2,4\},
\{1,3,4\}, \{2,3,4\}\}, \\
P_2 &=& \{\{1,4\}\}, P_3 = P_4 = \emptyset.
\end{eqnarray*}
By Theorem~\ref{thm:twist_gen_matroid} we have for $G*X$ with $X =
\{1,2,3\}$, depicted on the right-hand side of
Figure~\ref{fig:ex_graph_nullity_invar}, $\mathcal{P}_{G*X} =
(P'_0,P'_1,P'_2,P'_3,P'_4)$ where
\begin{eqnarray*}
P'_0 &=& \{\emptyset, \{2\}, \{4\}, \{1,2\}, \{1,3\}, \{1,2,3\},
\{1,2,4\}, \{1,3,4\}\},
\\
P'_1 &=& \{\{1\}, \{3\}, \{1,4\}, \{2,3\}, \{2,4\}, \{3,4\},
\{1,2,3,4\}\},
\\
P'_2 &=& \{\{2,3,4\}\}, P'_3 = P'_4 = \emptyset.
\end{eqnarray*}
We have $\|\mathcal{P}_G\| = \|\mathcal{P}_{G*X}\| = (8,7,1,0,0)$.
\end{Example}

\begin{Example} \label{ex:part_seq_overlap_graph}
In the context of Section~\ref{sec:background}, where matrix
$A$ an overlap graph $O_s$ for some double occurrence string $s$,
we have that $\|\mathcal{P}_{O_s}\|[i]$ is the number of partitions of the edges
of the 2-in, 2-out digraph $D$ corresponding to $s$ into closed
walks of $D$, such that the number of closed walks is precisely
$i+1$. The value $\|\mathcal{P}_{O_s}\|[0]$ is therefore the number
of Euler circuits in $D$.
\end{Example}

\section{Elementary Pivots on Graphs} \label{sec:pivots_graphs}
From now on we consider pivot on graphs (i.e., symmetric $V \times
V$-matrices over $\two$), and thus on all matrix computations will
be over $\two$. Hence for graph $G$, $\mathcal{M}_G = (V,D_G)$ is
the set system with $X \in D_G$ iff $\det(G\sub{X}) = 1$. Also, $G$
can be (re)constructed given $\mathcal{M}_G$. Indeed, $\{u\}$ is a
loop in $G$ iff $\{u\} \in D_G$, and $\{u,v\}$ is an edge in $G$ iff
$(\{u,v\} \in D_G) \xor ((\{u\} \in D_G) \wedge (\{v\} \in D_G))$,
see \cite[Property~3.1]{Bouchet_1991_67}. Therefore, the function
$\mathcal{M}_{(\cdot)}$ assigning to each graph $G$ the set system
$\mathcal{M}_G$ is an injective function from the family of graphs
to the family of set systems. It this way the family of graphs may
be regarded as a subclass of the family of set systems. Note that
$\mathcal{M}_G*X$ is defined for all $X \subseteq V$, while pivot on
graphs $G*X$ is defined only if $X \in \mathcal{M}_G$ (or
equivalently, $\emptyset \in \mathcal{M}_G*X$).

In this section we recall from \cite{Geelen97} that the pivot
operation on graphs can be defined as compositions of two graph
operations: local complementation and edge complementation.
%

The pivots $G*X$ where $X$ is a minimal element of $\mathcal{M}_G
\backslash \{\emptyset\}$ w.r.t. inclusion are called
\emph{elementary}. It is noted in \cite{Geelen97} that an elementary
pivot $X$ on graphs corresponds to either a loop, $X = \{u\} \in
E(G)$, or to an edge, $X = \{u,v\} \in E(G)$, where both vertices
$u$ and $v$ are non-loops. Thus for $Y \in \mathcal{M}_G$, if $G[Y]$
has elementary pivot $X_1$, then $Y \setminus X_1 = Y \xor X_1 \in
\mathcal{M}_{G*X_1}$. In this way, each $Y \in \mathcal{M}_G$ can be
partitioned $Y = X_1 \cup \cdots\cup X_n$ such that $G*Y = G*(X_1
\xor \cdots\xor X_n) = (\cdots(G*X_1)\cdots * X_n)$ is a composition
of elementary pivots. Consequently, a direct definition of the
elementary pivots on graphs $G$ is sufficient to define the
(general) pivot operation on graphs.

The elementary pivot $G*\{u\}$ on a loop $\{u\}$ is called
\emph{local complementation}. It is the graph obtained from $G$ by
complementing the edges in the neighbourhood $N_G(u) = \{ v \in V
\mid \{u,v\} \in E(G), u \not= v \}$ of $u$ in $G$: for each $v,w
\in N_G(u)$, $\{v,w\}\in E(G)$ iff $\{v,w\} \not\in E(G*\{u\})$, and
$\{v\}\in E(G)$ iff $\{v\} \not\in E(G*\{u\})$ (the case $v=w$). The
other edges are left unchanged.

The elementary pivot $G*\pair uv$ on an edge $\pair uv$ between
distinct non-loop vertices $u$ and $v$ is called \emph{edge
complementation}. For a vertex $x$ consider its closed neighbourhood
$N'_G(x)= N_G(x)\cup \{x\}$. The edge $\pair uv$ partitions the
vertices of $G$ connected to $u$ or $v$ into three sets $V_1 =
N'_G(u) \setminus N'_G(v)$, $V_2 = N'_G(v) \setminus N'_G(u)$, $V_3
= N'_G(u) \cap N'_G(v)$. Note that $u,v \in V_3$.

\begin{figure}[t]
\centerline{\unitlength 1.0mm
\begin{picture}(55,42)(0,1)
\drawccurve(02,28)(25,21)(48,32)(25,39)
\drawccurve(00,10)(10,00)(20,10)(10,20)
\drawccurve(30,10)(40,00)(50,10)(40,20)
\gasset{AHnb=0,Nw=1.5,Nh=1.5,Nframe=n,Nfill=y}
\gasset{ExtNL=y,NLdist=1.5,NLangle=90}
\put(10,02){\makebox(0,0)[cc]{$V_1$}}
\put(40,02){\makebox(0,0)[cc]{$V_2$}}
\put(25,36){\makebox(0,0)[cc]{$V_3$}}
  \node(u)(09,28){$u$}
  \node(v)(20,30){$v$}
  \node(uu)(29,32){}
  \node(vv)(41,28){}
  \node(u1)(7,14){}
  \node(u2)(14,7){}
  \node(v1)(38,7){}
  \node(v2)(43,14){}
  \drawedge(u,v){}
  \drawedge(u,u1){}
  \drawedge(u,u2){}
  \drawedge(v,v1){}
  \drawedge(v,v2){}
  \drawedge(u1,v2){}
  \drawedge(u2,v1){}
  \drawedge[dash={1}0](v1,v2){}
  \drawedge[dash={1}0](u1,u2){}
  \drawedge[dash={1}0](uu,vv){}
  \drawedge(uu,u1){}
  \drawedge(vv,u2){}
  \drawedge(uu,v1){}
  \drawedge(vv,v2){}
\end{picture}
\begin{picture}(55,42)(0,1)
\drawccurve(02,28)(25,21)(48,32)(25,39)
\drawccurve(00,10)(10,00)(20,10)(10,20)
\drawccurve(30,10)(40,00)(50,10)(40,20)
\gasset{AHnb=0,Nw=1.5,Nh=1.5,Nframe=n,Nfill=y}
\gasset{ExtNL=y,NLdist=1.5,NLangle=90}
\put(10,02){\makebox(0,0)[cc]{$V_1$}}
\put(40,02){\makebox(0,0)[cc]{$V_2$}}
\put(25,36){\makebox(0,0)[cc]{$V_3$}}
  \node(u)(09,28){$u$}
  \node(v)(20,30){$v$}
  \node(uu)(29,32){}
  \node(vv)(41,28){}
  \node(u1)(7,14){}
  \node(u2)(14,7){}
  \node(v1)(38,7){}
  \node(v2)(43,14){}
  \drawedge(u,v){}
  \drawedge(v,u1){}
  \drawedge(v,u2){}
  \drawedge(u,v1){}
  \drawedge(u,v2){}
  \drawedge(u1,v1){}
  \drawedge(u2,v2){}
  \drawedge[dash={1}0](v1,v2){}
  \drawedge[dash={1}0](u1,u2){}
  \drawedge[dash={1}0](uu,vv){}
  \drawedge(uu,u2){}
  \drawedge(vv,u1){}
  \drawedge(uu,v2){}
  \drawedge(vv,v1){}
\end{picture}%
} \caption{Pivoting on an edge $\pair uv$ in a graph with both $u$
and $v$ non loops. Connection $\pair xy$ is toggled iff $x\in V_i$
and $y\in V_j$ with $i\neq j$. Note $u$ and $v$ are connected to all
vertices in $V_3$, these edges are omitted in the diagram. The
operation does not affect edges adjacent to vertices outside the
sets $V_1,V_2,V_3$, nor does it change any of the loops.}%
\label{fig:pivot}
\end{figure}
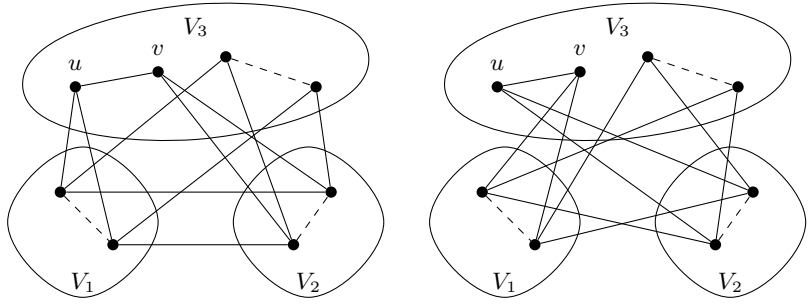

The graph  $G*\pair uv$ is constructed by ``toggling'' all edges
between different $V_i$ and $V_j$: for $\pair xy$ with $x\in V_i$,
$y\in V_j$ and $i\neq j$: $\pair xy \in E(G)$ iff $\pair xy \notin
E(G\sub{\{u,v\}})$, see Figure~\ref{fig:pivot}. The other edges
remain unchanged. Note that, as a result of this operation, the
neighbours of $u$ and $v$ are interchanged.

\renewcommand{\pivotorbitmacro}{
  \gasset{AHLength=0,Nw=8,Nh=8,Nframe=y,Nfill=n}
  \node(p)(30,30){$p$}
  \node(q)(0,0){$q$}
  \node(r)(30,0){$r$}
}

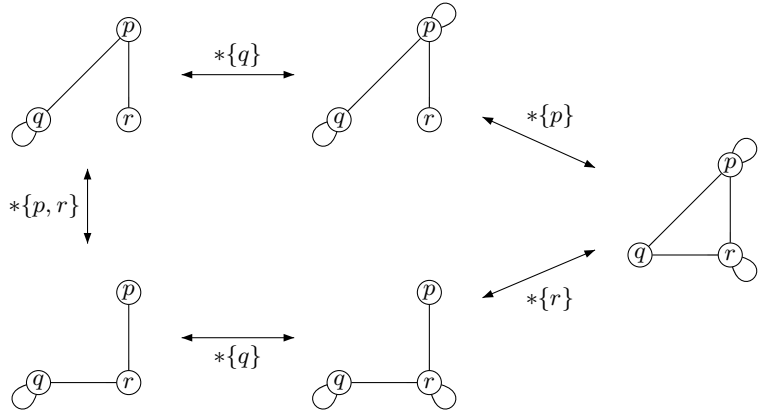
\begin{figure}[t]
\unitlength 1mm%
\begin{center}
%
\begin{picture}(100,48)(-10,-05)
\node[Nw=25,Nh=25,Nframe=n](I)(00,35){%
  \unitlength0.4mm
  \begin{picture}(30,30)
  \pivotorbitmacro
  \drawedge(p,q){}
  \drawedge(p,r){}
  \drawloop[loopangle=-135,loopdiam=7](q){}
  \end{picture}
}
\node[Nw=25,Nh=25,Nframe=n](II)(40,35){%
  \unitlength0.4mm
  \begin{picture}(30,30)
  \pivotorbitmacro
  \drawedge(p,q){}
  \drawedge(p,r){}
  \drawloop[loopangle=45,loopdiam=7](p){}
  \drawloop[loopangle=-135,loopdiam=7](q){}
\end{picture}
}
\node[Nw=25,Nh=25,Nframe=n](III)(80,17){%
  \unitlength0.4mm
  \begin{picture}(30,30)
  \pivotorbitmacro
  \drawedge(p,q){}
  \drawedge(p,r){}
  \drawedge(q,r){}
  \drawloop[loopangle=45,loopdiam=7](p){}
  \drawloop[loopangle=-45,loopdiam=7](r){}
\end{picture}
}
\node[Nw=25,Nh=25,Nframe=n](IV)(40,00){%
  \unitlength0.4mm
  \begin{picture}(30,30)
  \pivotorbitmacro
  \drawedge(p,r){}
  \drawedge(q,r){}
  \drawloop[loopangle=-135,loopdiam=7](q){}
  \drawloop[loopangle=-45,loopdiam=7](r){}
\end{picture}
}
\node[Nw=25,Nh=25,Nframe=n](V)(00,00){%
  \unitlength0.4mm
  \begin{picture}(30,30)
  \pivotorbitmacro
  \drawedge(p,r){}
  \drawedge(q,r){}
  \drawloop[loopangle=-135,loopdiam=7](q){}
\end{picture}
}
  \drawedge[AHnb=1,ATnb=1](I,II){$*\{q\}$}
  \drawedge[AHnb=1,ATnb=1](II,III){$*\{p\}$}
  \drawedge[AHnb=1,ATnb=1](III,IV){$*\{r\}$}
  \drawedge[AHnb=1,ATnb=1](IV,V){$*\{q\}$}
  \drawedge[AHnb=1,ATnb=1](V,I){$*\{p,r\}$}
\end{picture}
\end{center}
\caption{The orbit of a graph under pivot. Only the elementary
pivots are shown.}\label{fig:pivot_space}
\end{figure}

\begin{Example}
Figure~\ref{fig:pivot_space} depicts an orbit of graphs under pivot.
The figure also shows the applicable elementary pivots (i.e., local
and edge complementation) of the graphs within the orbit.
\end{Example}

Interestingly, in many contexts, principal pivot transform
originally appeared in disguise. For example, PPT was recognized in
\cite{GlantzP06} as the operation underlying the recursive
definition of the interlace polynomial, introduced in
\cite{Arratia_InterlaceP_SODA}. We will consider the interlace
polynomial in the next section. Also, e.g., the graph model defined
in \cite{Equiv_String_Graph_2} within the formal theory of
(intramolecular) gene assembly in ciliates turns out to be exactly
the elementary pivots, as noted in \cite{BHH/PivotsDetPM/09}.
Furthermore, the proof of the result from
\cite{DBLP:journals/jct/CohnL72}, connecting nullity to the number
of cycles in permutations, as mentioned in
Section~\ref{sec:background}, implicitly uses the Schur complement
(which is an essential part of PPT).

\section{The Interlace Polynomial}
The interlace polynomial is a graph polynomial introduced in
\cite{Arratia_InterlaceP_SODA,Arratia2004199}. We follow the
terminology of \cite{ArratiaBS04}. The single-variable interlace
polynomial (simply called interlace polynomial in
\cite{Arratia2004199}) for a graph $G$ (with possibly loops) is
defined by
$$
q(G) = \sum_{S\subseteq V} (y-1)^{n(G\sub{S})}.
$$
It is is shown in \cite{ArratiaBS04} that the interlace polynomial
fulfills an interesting recursive relation, cf.
Proposition~\ref{prop:arratia_recursion} below, which involves local
and edge complementation. As we consider here its generalization,
principal pivot transform, it makes sense now to define the
interlace polynomial for $V \times V$-matrices (over some arbitrary
field) in general. Therefore, we define the \emph{interlace
polynomial} for $V \times V$-matrix $A$ as
$$
q(A) = \sum_{S\subseteq V} (y-1)^{n(A\sub{S})}.
$$
We may (without loss of information) change variables $y := y-1$ in
the definition of the interlace polynomial to obtain
$$
q'(A) = \sum_{S\subseteq V} y^{n(A\sub{S})}.
$$
As $q(A)$ (and $q'(A)$) deals with nullity values for (square)
matrices in general, one can argue that the \emph{nullity
polynomial} is a more appropriate name for these polynomials.

We see that the coefficient $a_i$ of term $a_i y^i$ of $q'(A)$ is
equal to $\|\mathcal{P}_{A}\|[i]$ (the element of
$\|\mathcal{P}_{A}\|$ corresponding to $i$) for all $i \in
\{0,\ldots,n\}$. Therefore, we have for matrices $A$ and $A'$, $q(A)
= q(A')$ iff $q'(A) = q'(A')$ iff $\|\mathcal{P}_{A}\| =
\|\mathcal{P}_{A'}\|$.

\begin{Example} \label{ex:int_poly_overlap_graph}
Let $O_s$ be the overlap graph for some double occurrence string
$s$, and let $a_i$ be the coefficient $a_i$ of term $a_i y^i$ of
$q'(O_s)$. We have, see Example~\ref{ex:part_seq_overlap_graph},
that $a_i$ is equal to the number of partitions of the edges of the
2-in, 2-out digraph $D$ corresponding to $s$ into closed walks of
$D$, such that the number of closed walks is precisely $i+1$. More
specifically, $a_0$ is the number of Euler circuits in $D$. The
interlace polynomial is originally motivated by the computation of
these coefficients $a_i$ of 2-in, 2-out digraphs, see
\cite{Arratia2004199}.
\end{Example}

It is shown in \cite{Arratia2004199} that the interlace polynomial
is invariant under edge complementation. By
Theorem~\ref{thm:twist_gen_matroid} we see directly that this holds
for pivot in general: $\|\mathcal{P}_{A*X}\| = \|\mathcal{P}_{A}\|$
and equivalently $q(A*X) = q(A)$.

Furthermore, we show that $q(A)$ fulfills the following recursive
relation.
\begin{Theorem} \label{thm:disjoint_sum}
Let $A$ be a $V \times V$-matrix (over some field), let $X \subseteq
V$ with $A[X]$ nonsingular, and let $u \in X$. We have $q(A) =
q(A\vertexrem u) + q(A*X\vertexrem u)$.
\end{Theorem}
\begin{Proof}
Let $\mathcal{P}_A = (P_0,P_1,\ldots,P_n)$. Since $X$ is nonempty
and $A[X]$ is nonsingular, $P_n = \emptyset$. Let $R =
(P_0,P_1,\ldots,P_{n-1})$. Let $Z \in P_i$ for $i \in
\{0,1,\ldots,n-1\}$. We have $Z \subseteq V$ \emph{does not} appear
in $\mathcal{P}_{A\vertexrem u}$ iff $u \in Z$ iff $u \not\in Z \xor
X$ iff $Z \xor X$ \emph{does} appear in $\mathcal{P}_{A*X\vertexrem
u}$. Hence $\|R\| = \|\mathcal{P}_{A\vertexrem u}\| +
\|\mathcal{P}_{A*X\vertexrem u}\|$ (point-wise addition of the two
vectors), and the statement holds.
\end{Proof}


The recursive relation for the single-variable interlace polynomial
in \cite{ArratiaBS04} is now easily obtained from
Theorem~\ref{thm:disjoint_sum} by restricting to the case of
elementary pivots on graphs.\footnote{We use here the fact observed
in \cite{GlantzP06} that the operations in the recursive relations
of \cite{ArratiaBS04} are exactly the elementary pivots of
Section~\ref{sec:pivots_graphs}, assuming that the neighbours of $u$
and $v$ are interchanged after applying the ``pivot'' operation of
\cite{ArratiaBS04} on edge $\{u,v\}$.}
\begin{Proposition}[\cite{ArratiaBS04}]
\label{prop:arratia_recursion} Let $G$ be a graph. Then $q(G)$
fulfills the following conditions.
\begin{enumerate}
\item $q(G) = q(G\vertexrem u) + q(G*\{u,v\}\vertexrem u)$ for
edge $\{u,v\}$ in $G$ where both $u$ and $v$ do not have a loop,
\item $q(G) = q(G\vertexrem u) + q(G*\{u\}\vertexrem u)$ if
$u$ has a loop in $G$, and
\item $q(G) = y^n$ if $G$ is a
discrete graph with $n$ vertices.
\end{enumerate}
\end{Proposition}
\begin{Proof}
Conditions (1) and (2) follow from Theorem~\ref{thm:disjoint_sum}
where $A$ is a graph $G$, and $X$ is an elementary pivot
(i.e., $X = \{u\}$ is a loop in $G$ or $X = \{u,v\}$ is an edge in
$G$ where both $u$ and $v$ do not
have a loop, see Section~\ref{sec:pivots_graphs}). Finally, if $G$
is a discrete graph with $n$ vertices, then, for all $Y \subseteq
V$, $Y \in P_{|Y|}$. Consequently, $|P_i| = {n \choose i}$. Thus,
$q'(G) = (y+1)^n$ and therefore $q(G) = y^n$.
\end{Proof}

\section{Discussion}
We have shown that the pivot operator $*X$ on matrices $A$ and the
symmetric difference operator $\xor X$ on sets $Y$ have an
equivalent effect w.r.t. the nullity value of the principal
submatrices $A\sub{Y}$ of $A$. This nullity invariant may be
described in terms of partition sequences $\mathcal{P}_A$, where the
sets $Y \subseteq V$ are arranged according to the nullity value of
$A\sub{Y}$. We notice that interlace polynomial of a graph $G$
corresponds to the norm $\|\mathcal{P}_G\|$ of the partition
sequence of $G$ (where $G$ is considered as a matrix). Hence we
(may) naturally consider interlace polynomials for square matrices
in general, and obtain a recursive relation for these generalized
interlace polynomials. In this way, we simplify the proof of the
(original) recursive relation for interlace polynomials of graphs.

\subsection*{Acknowledgements} R. Brijder is supported by the Netherlands Organization for
Scientific Research (NWO), project ``Annotated graph mining''.

\bibliography{../geneassembly}

\end{document}